\documentclass[epj,final]{svjour}
\usepackage{psfig}

\begin{document}
\title{K$^+$ production in proton-nucleus reactions and the role of
momentum-dependent potentials}
\author{Z.
Rudy\inst{1}            \and W. Cassing\inst{2}   \and L. Jarczyk\inst{1}         \and B.
Kamys\inst{1}    \and P. Kulessa\inst{1,3}
}                     
\institute{ M. Smoluchowski Institute of Physics, Jagellonian University,
PL-30059 Cracow, Poland \and
Institut f\"ur Theoretische Physik, Justus Liebig Universit\"at
Giessen, D-35392 Giessen, Germany    \and
Institut f\"ur Kernphysik, Forschungszentrum J\"ulich,
D-52425 J\"ulich, Germany }
\date{Received: date / Revised version: date}
\abstract{ The production of $K^+$  mesons in proton-nucleus
collisions from 1.0 to 2.5 GeV  is analyzed with respect to
one-step nucleon-nucleon
 $(NN\rightarrow N Y K^+$)
 and two-step $\Delta$-nucleon $(\Delta N \rightarrow  K^+ Y N$) or
 pion-nucleon  $(\pi N \rightarrow  K^+ Y $)
 production channels on the basis of a coupled-channel transport
 approach (CBUU) including the kaon final state interactions.
The influence of momentum-dependent potentials for the nucleon,
hyperon and kaon in the final state are studied as well as the
importance of $K^+$ elastic rescattering in the target nucleus.
The transport calculations are compared to the experimental $K^+$
spectra taken at LBL Berkeley, SATURNE, CELSIUS, GSI and
COSY-J\"ulich. It is found that the momentum-dependent baryon
potentials effect the excitation function of the $K^+$ cross
section; at low bombarding energies of $\sim $ 1.0 GeV the
attractive baryon potentials in the final state lead to a relative
enhancement of the kaon yield whereas the net repulsive potential
at bombarding energies $\sim$ 2 GeV causes a decrease of the $K^+$
cross section. Furthermore it is pointed out, that especially the
$K^+$ spectra  at low momenta (or kinetic energy $T_K$) allow to
determine the in-medium $K^+$ potential almost model independently
due to a relative shift of the $K^+$ spectra in kinetic energy
that arises from the acceleration of the kaons when propagating
out of the nuclear medium to free space, i.e. converting potential
energy to kinetic energy of the free kaon.
} 

\maketitle

\section{Introduction}
The production of  mesons heavier than pions in $p+A$ reactions at
bombarding energies far below and close to the free
nucleon-nucleon threshold is of specific interest
\cite{1}-\cite{cass95} as one hopes to learn either about
cooperative nuclear phenomena and/or about high momentum
components of the nuclear many-body wave function that arise from
nucleon-nucleon correlations. Especially $K^+$ mesons have been
considered as promising hadronic probes \cite{5,12} due to the
rather moderate final state interaction, which is a consequence of
 strangeness conservation and the fact that there are no
baryon resonances with anti-strange quarks in nuclei.
Anti-hyperons, furthermore, have a much larger production
threshold and annihilate very fast in nuclei. On the other hand,
the kaon properties might change in the nuclear medium
\cite{Kaplan} such that conclusions on cooperative nuclear
phenomena require a precise understanding of the (anti-) kaon
potentials at finite nuclear density.

Experiments on $K^\pm$ production from nucleus-nucleus collisions
at SIS energies of 1-- 2 A$\cdot$ GeV have shown that in-medium
properties of the kaons are seen in the collective flow pattern of
$K^+$ mesons both, in-plane and out-of-plane, as well as in the
abundancy  of antikaons \cite{cass99,Li2001}. Thus in-medium
modifications of the mesons have become a topic of substantial
interest in the last decade triggered in part by the early
suggestion of Brown and Rho \cite{BR}, that the modifications of
hadron masses should scale with the scalar quark condensate
$<q\bar{q}>$ at finite baryon density.

As demonstrated in the pioneering work of Kaplan and Nelson
\cite{Kaplan} kaons and antikaons couple attractively to the
scalar nucleon density with a strength proportional to the $KN-
\Sigma$ constant,
\begin{equation} \label{sigmat}
\Sigma_{KN} = \frac{1}{2} (m^0_u + m^0_s) \ <N|\bar{u}u +
\bar{s}s|N>,
\end{equation}
which is not well known at present and may vary from 270 to 450
MeV. In (\ref{sigmat}) $m^0_u$ and $m^0_s$ denote the bare masses
for the light $u$- and strange $s$-quark while the expression in
the brackets is the expectation value of the scalar light and
strange quark condensate for the nucleon \cite{cass99}.
Furthermore, a vector coupling to the quark 4-current -- for
vanishing spatial components -- leads to a repulsive potential
term for the kaons; on the other hand this (Weinberg-Tomozawa)
term is attractive for the antikaons.

In chiral effective theories the dispersion relation for kaons and
antikaons in the nuclear medium -- for low momenta --  can be
written as \cite{nelson}
\begin{eqnarray}
&&\omega_K(\rho_N,{\bf p}) = + \frac{3}{8} \frac{\rho_N}{f_\pi^2}
\label{nelson1}\\ && + \left[ {\bf p}^2 + m_K^2 \left( 1 -
\frac{\Sigma_{KN}}{f_\pi^2 m_K^2} \rho_s + \left( \frac{3\rho_N}{8
f_\pi^2 m_K} \right)^2 \right) \right]^{1/2} , \nonumber \\
&&\omega_{\bar K} (\rho_N,{\bf p}) =  - \frac{3}{8}
\frac{\rho_N}{f_\pi^2}\label{nelson2} \\ &&+ \left[ {\bf p}^2 +
m_K^2 \left( 1 - \frac{\Sigma_{KN}}{f_\pi^2 m_K^2} \rho_s + \left(
\frac{3\rho_N}{8 f_\pi^2 m_K} \right)^2 \right) \right]^{1/2}.
\nonumber\end{eqnarray} In equations (2), (3)  $m_K$ denotes the
bare kaon mass, $f_\pi \approx$ 93 MeV is the pion decay constant,
while $\rho_s$ and $\rho_N$ stand for the scalar and vector
nucleon densities, respectively. As shown in Ref.
\cite{Schaffner}, for $\Sigma_{KN}$= 450 MeV  one ends up with an
effective kaon potential which is close to zero below ordinary
nuclear matter density $\rho_0$ and becomes  more repulsive above
$\rho_0$. On the other hand, using $\Sigma_{KN}$= 270 MeV a
repulsive kaon potential of $\approx$ 25 MeV at normal nuclear
matter density is obtained. Note, that when extrapolating
(\ref{nelson2}) to 3$\rho_0$ and above, the antikaon mass becomes
very light. Thus antikaon condensates might occur at high baryon
density which, furthermore, are of great interest in the
astrophysical context \cite{GB1,muto,sahu01}.

However, the actual kaon and antikaon self energies (or
potentials) are quite a matter of debate -- due to higher order
terms in the chiral expansion -- especially for the antikaon
\cite{Gal,Lutz,Ramos} and the momentum-dependence of their self
energies  is widely unknown (except for a dispersion analysis in
Ref. \cite{Sib98}) since most Lagrangian models restrict to
$s$-wave interactions or only include additional $p$-waves. It is
thus mandatory to perform experimental studies of the (anti-) kaon
properties under well controlled conditions, e.g. in
proton-nucleus reactions, where one probes the (anti-) kaon self
energies  at normal nuclear matter density $\rho_0
\approx$ 0.16 fm$^{-3}$ and below. Furthermore, by gating on kaon momenta in
the laboratory, one might be able to obtain information on the
momentum dependence of the self energies, too.

 $K^+$ production in $p+A$ collisions at subthreshold energies
has been observed experimentally more than a decade ago by Koptev
et al. \cite{12} at bombarding energies from 0.8--1.0 GeV.
However, only total $K^+$ yields could be extracted at that time.
Nevertheless, the target-mass dependence of the $K^+$ yield ($\sim
A$) suggested the dominance of two-step reactions with an
intermediate pion or $\Delta$. Detailed folding-model calculations
in Refs. \cite{15,cass95} essentially came to the same conclusion.
First differential $K^+$ spectra from $p+NaF$ and $p+Pb$ reactions
from the LBL Berkeley had been performed at $T_{lab}$ = 2.1 GeV
\cite{schnetzer}, i.e. far above the $NN$ threshold of 1.58 GeV in
free space. Only in more recent years differential $K^+$ spectra
have been measured down to 1.2 GeV for $^{12}C$ targets at 40 $^0$
\cite{debowski} (at SATURNE) or 90 $^0$ in the laboratory
\cite{badala} (at CELSIUS). Unfortunately, the different
experiments have no overlap in acceptance and the interpretation
of the data, if compatible at all, remains vague \cite{Markus}.
First data on the full momentum distribution at forward angles
have been presented very recently by the ANKE Collaboration at
COSY-J\"ulich \cite{Barsov} for $K^+$ mesons from $p+^{12}C$
reactions at 1.0 GeV \cite{ANKE}. These data show a kinematical
focussing of the spectra at finite momentum of $\approx$ 350
MeV/c, which appears incompatible with the cross section from Ref.
\cite{badala}. Thus a systematic comparison of all data is
urgently needed within an adequate theoretical approach that
allows to compare the kinematically restricted data on the same
footing.

In this study we use the coupled-channel (CBUU) transport model
that has first been developed in Ref. \cite{Wolf} for the
description of nucleus-nucleus collisions and later on employed
for the simulation of pion- and proton-nucleus reactions
\cite{Sib98,zibi95,cass98}, too. For applications to $K^\pm$
production in nucleus-nucleus collisions at SIS energies we refer
the reader to Ref. \cite{cass97}. In this model the effects of
momentum-dependent self energies for all hadrons can be studied
explicitly as well as their production and propagation in the
nuclear medium.

The paper is organized as follows: We briefly recapitulate the
ingredients of the CBUU model in Section 2, present the extensions
performed in this study and investigate in particular the effects
from $K^+$ rescattering. In section 3 we compare the transport
calculations  to the experimental spectra available from different
laboratories and explore the sensitivity of the $K^+$ spectra to
the momentum-dependent potentials employed. A summary and
discussion of open problems concludes this paper in Section 4.

\section{Ingredients of the transport model}

In this work we perform our analysis along the line of the
CBUU\footnote{Coupled-Channel Boltzmann-Uehling-Uhlenbeck}
 approach~\cite{Wolf} which is based on a
coupled set of  transport equations for the phase-space
distributions $f_{h} (x,p)$ of hadron $h$, i.e.
\cite{weber,Ehehalt}
\begin{eqnarray}
&& \left( \Pi_{\mu}-\Pi_{\nu}\partial_{\mu}^p
U_{h}^{\nu} -M_{h}^*\partial^p_{\mu} U_{h}^{S}
\right) \partial_x^{\mu} f_{h}(x,p) \label{g24} \\
&&+ \left( \Pi_{\nu} \partial^x_{\mu}
U^{\nu}_{h}+ M^*_{h} \partial^x_{\mu}U^{S}_{h}\right)
\partial^{\mu}_p  f_{h}(x,p)   \nonumber\\
&&= \sum_{h_2 h_3 h_4\ldots} \int d2 d3 d4 \ldots
 [G^{\dagger}G]_{12\to 34\ldots} \nonumber \\
&&\phantom{a}\hspace*{2cm}\times\delta^4(\Pi +\Pi_2-\Pi_3-\Pi_4 \ldots ) \nonumber\\
&& \times \left\{ f_{h_3}(x,p_3)f_{h_4}(x,p_4)\bar{f}_{h}(x,p)
\bar{f}_{h_2}(x,p_2) \right.\nonumber \\
&& - \left. f_{h}(x,p)f_{h_2}(x,p_2)\bar{f}_{h_3}(x,p_3) \bar{f}_{h_4}(x,p_4)
\right\} \ldots\ \ .
\end{eqnarray}
In Eq.~(\ref{g24}) $U_{h}^{S}(x,p)$ and $U_{h}^{\mu}(x,p)$ denote
the real part of the scalar and vector hadron selfenergies,
respectively, while $[G^\dagger G]_{12\to 34\ldots} \delta^4 (\Pi
+\Pi_2-\Pi_3-\Pi_4\ldots )$ is the 'transition rate' for the
process $1+2\to 3+4+\ldots$ which is taken to be on-shell in the
semiclassical limit adopted. The hadron quasi-particle properties
in (\ref{g24}) are defined via the mass-shell constraint
\begin{equation}   \label{g25}
\delta (\Pi_{\mu}\Pi^{\mu}-M_{h}^{*2} ) \ \ ,
\end{equation}
with effective masses and momenta (in local Thomas-Fermi
approximation) given by \cite{weber}
\begin{eqnarray}\label{g26}
M_{h}^* (x,p)&=&M_h + U_h^{{S}}(x,p) \nonumber \\ \Pi^{\mu}
(x,p)&=&p^{\mu}-U^{\mu}_h (x,p)\ \ ,
\end{eqnarray}
while the phase-space factors
\begin{equation}
\bar{f}_{h} (x,p)=1 \pm f_{{h}} (x,p)
\end{equation}
are responsible for fermion Pauli-blocking or Bose enhancement,
respectively, depending on the type of hadron in the final/initial
channel. The dots in Eq.~(5) stand for further contributions to
the collision term with more than two hadrons in the final/initial
channels (cf. Ref. \cite{Cassing01}). The transport approach
(\ref{g24}) is fully specified by $U_{h}^{S}(x,p)$ and
$U_{h}^{\mu}(x,p)$ $(\mu =0,1,2,3)$, which determine the
mean-field propagation of the hadrons, and by the transition rates
$G^\dagger G\,\delta^4 (\ldots )$ in the collision term (5), that
describes the scattering and hadron production/absorption rates.

\begin{figure}[h]
\centerline{\psfig{figure=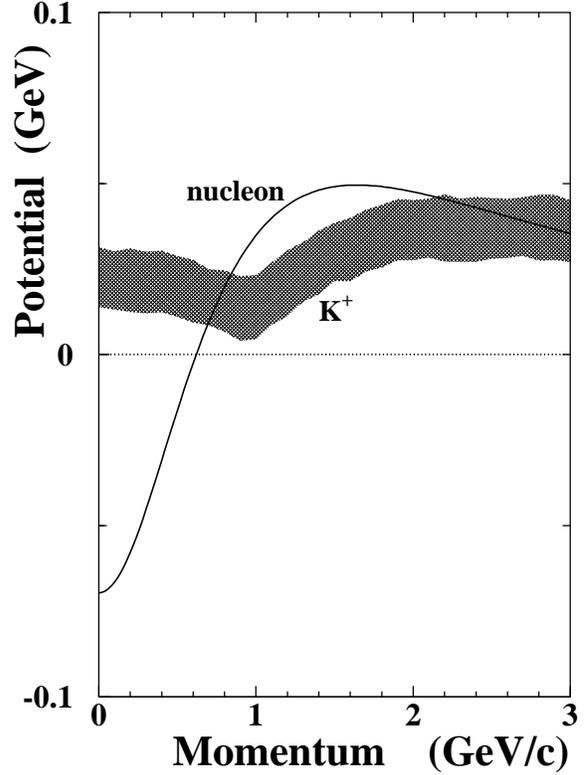,width=7.5cm}}
 \caption{The nucleon potential (solid line)  at density
$\rho_0$ as a function of momentum relative to the nuclear matter
at rest as used in the CBUU transport approach. The hatched area
denotes the  $K^+$ potential at $\rho_0$ (within the
uncertainties) from the dispersion analysis in Ref. \cite{Sib98}.}
 \label{bild1}
\end{figure}

The scalar and vector mean fields $U_{h}^{S}$ and $U^\mu_{h}$ for
nucleons are modeled as in  Ref.~\cite{Ehehalt}, however, slightly
modified in line with Ref. \cite{excita}. In Fig. \ref{bild1} the
real part of the Schroedinger equivalent potential (SEP) for
nucleons,
\begin{equation}
U_{SEP}(p) = \Pi^0({\bf p}) - \sqrt{{\bf p}^2 + M_0^2}, \label{Ehg23}
\end{equation} is shown at density $\rho_0$ (full line) as a
function of the momentum  $p$ relative to the nuclear matter at
rest. Whereas we see a net attraction for momenta $p \leq$
0.6~GeV/c, the nucleon potential becomes repulsive for higher
momenta and  reaches a maximum repulsion at $p \approx$ 1.5 GeV/c.
We mention that at density $\rho_0$ the Schroedinger equivalent
potential $U_{SEP}$ compares well with the potential from the data
analysis of Hama et al. \cite{Hama} as well as Dirac-Brueckner
computations from \cite{Mal} up to a kinetic energy $E_{kin}$ of 1
GeV \cite{Ehehalt}.

Apart from the nuclear potentials each charged hadron additionally
moves in the background of the Coulomb potential that is generated
by the charged hadrons themselves. In case of proton-nucleus
reactions -- with the nucleus at rest -- this essentially amounts
to a Coulomb acceleration in the final state for positively
charged hadrons. Note, that for heavy nuclei like $Pb$ or $Au$ the
Coulomb potential in the nuclear interior is about + 20 MeV, i.e.
of the same order of magnitude as the 'expected' repulsive $K^+$
nuclear potential.

 The hyperon mean fields, furthermore, are assumed to be 2/3
of the nucleon potentials. In the present approach, apart from
nucleons, $\Delta, N(1440)$, $N(1535)$, $\Lambda, \Sigma$ with
their isospin degrees of freedom, we propagate explicitly pions,
$K^+, K^-$, and $\eta$'s and assume that the pions as Goldstone
bosons do not change their properties in the medium; we also
discard self energies for the $\eta$-mesons which have a minor
effect on the kaon dynamics.  The kaon and antikaon potentials,
however, have to be specified explicitly.

\subsection{$K^+, K^-$ self energies}

Apart from (\ref{nelson1}), (\ref{nelson2}) there is a couple of
models for the kaon and antikaon
selfenergies~\cite{Kaplan,GB1,waas}, which differ in the actual
magnitude, however, agree on the relative signs for kaons and
antikaons. Thus in line with the kaon-nucleon scattering amplitude
the $K^+$ potential should be slightly repulsive at finite baryon
density whereas the antikaon should see an attractive potential in
the nuclear medium. Without going into a detailed discussion of
the various models (cf. Ref. \cite{Schaffner} and figures therein)
we adopt the more practical point of view, that the actual $K^+$
and $K^-$ self energies are unknown and as a guide for our
analysis use a linear extrapolation of the form,
\begin{equation}
\label{kmass} m^*_K(\rho_B,p) = m_K^0 \left(1 + \alpha(p)
\frac{\rho_B}{\rho_0}\right),
\end{equation}
with $\alpha(p)$ denoting a momentum-dependent coefficient.
Following the dispersion analysis of Sibirtsev et al. \cite{Sib98}
the coefficients $\alpha(p)$ can be modelled in line with the
$K^\pm$ potentials from Fig. 9 of \cite{Sib98}; the resulting kaon
potential $U_{K^+}$ at density $\rho_0$ is shown in Fig.
\ref{bild1} in terms of the hatched area and remains repulsive for
all momenta considered. In the momentum range of interest in this
study, i.e. from 0.1 -- 1.0 GeV/c, the $K^+$ potential at density
$\rho_0$ may be approximately represented by a constant of
$U_{K^+} \approx 20$ MeV taking into account the relative
uncertainty of $\pm$ 10 MeV from the analysis in Ref.
\cite{Sib98}. Since the antikaon dynamics has been investigated in
Ref. \cite{Sib98} for $p+A$ reactions, we skip a further
description of the actual implementation of the $K^-$ potential.

\subsection{Perturbative treatment of strangeness production}
The calculation of 'subthreshold' particle production has to be
treated perturbatively in the energy regime of interest due
to the small cross sections involved. Since we work within the
parallel ensemble algorithm, each parallel run of the transport
calculation can be considered approximately as an individual
reaction event, where binary reactions in the entrance channel at
given invariant energy $\sqrt{s}$ lead to final states with 2
(e.g. $K^+ Y$ in $\pi B$ channels), 3 (e.g. for $K^+ YN$ channels
in $BB$ collisions) or 4 particles (e.g. $K\bar{K}NN$ in $BB$
collisions) with a relative weight $P_i$ for each event $i$ which
is defined by the ratio of the  production cross section to the
total hadron-hadron cross section\footnote{The actual final states
are chosen  according to the 2, 3, or 4-body phase space.}. We
thus dynamically gate on all events where a $K^+ Y$ or $K^+ K^-$
pair is produced initially.  Each strange hadron production event
$i$ is represented by  $\sim 10^3$ testparticles for the final
strange hadron $j$ with individual weight $W_j^i$ such that the
sum of the weights $W_j^i$ over $j$ reproduces the individual
production probability $P_i$ and the distribution in momenta
(multiplied with the $NN$ or $\pi N$ cross section) describes the
differential production cross section. Then the 'perturbative'
hadrons are propagated according to the Hamilton equations of
motion including the potentials. Elastic and inelastic reactions
with pions, $\eta$'s or nonstrange baryons are computed in the
standard way~\cite{Wolf}.

The final differential cross sections are obtained by multiplying
each testparticle weight $W_j^i$ by the total inelastic $pA$ cross section
and gating on the experimental acceptance of the different
detectors. In this way one achieves a realistic simulation of the
strangeness production, propagation and reabsorption during the
proton-nucleus collision with sufficient statistics to allow for
selective cuts also at the low bombarding energy of 1.0 GeV,
where the total $K^+$ cross section is in the order of 1 $\mu b$
(for $Au$) or below (for $^{12}C$).

\subsection{Elementary cross sections}
For the present study the production of pions by $pN$ collisions
in $p+A$ reactions as well the total kaon cross sections in $pN$
and $\pi N$ collisions are of relevance. The pion production cross
section from $NN$ interactions is based on the parametrization of
the experimental data by Ver West and Arndt \cite{21} and
implemented in the transport model as described in Ref.
\cite{Wolf}. The cross sections for the channels $\pi N
\rightarrow K Y$, where $Y$ stands for a hyperon $\Lambda,
\Sigma$, are taken from the analysis of Huang et al. \cite{huang}
and essentially correspond to the experimental data for the
different $\pi N$ channels in vacuum (or 'free' space). Note, that
in addition to our early studies in \cite{15,cass95} the channels
with a $\Sigma$ hyperon are  taken into account. All cross
sections are reparametrized as a function of the invariant energy
above threshold $\sqrt{s}-\sqrt{s_0}$ \cite{cass99}, where
$\sqrt{s_0}$ denotes the threshold for the individual channel
given by the sum of the hadron masses in the final state of the
reaction.

\begin{figure}[h]
\centerline{\psfig{figure=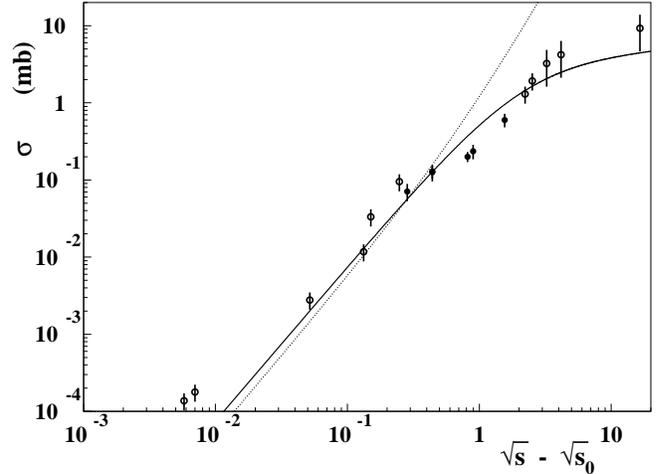,width=8.5cm,angle=270}}
 \caption{Comparison
of our parametrization for the inclusive $K^+$ production cross
section in $pp$ reactions (solid line) with the data from Refs.
\cite{COSY11,LB}. The dotted line corresponds to the
parametrization from Zwermann \cite{zwermann} used earlier in
Refs. \cite{15,cass95}. }
 \label{bild2}
\end{figure}

Whereas the production cross section of kaons from $pN$ collisions
close to threshold has been essentially unknown about a decade
ago, the ambiguity in this cross section has been resolved
experimentally  by now \cite{COSY11} and more adequate
parametrizations of the cross sections can be employed. The
experimental data from Refs. \cite{COSY11,LB} on the $pp
\rightarrow K^+ + X$ reactions are displayed in Fig. \ref{bild2}
in comparison to the current approximation (solid line) and the
estimate from Zwermann \cite{zwermann} (dotted line) used earlier
\cite{15,cass95}. Thus the problem of 'free' cross sections for
$pN$ collisions is sufficiently under control. For $\Delta N$
collisions, however, no experimental data are available. We use
the same cross section as a function of the invariant energy
$\sqrt{s}$ as in the $pN$ case keeping in mind this basic
uncertainty.

The question arises, furthermore, about the in-medium production
cross sections - essentially at density $\rho_0$ - if potentials
or self energies are involved. Here we employ the assumption that
the production matrix element squared $|M|^2$ does not change in
the medium and the change of cross section can be described by a
change of the available phase space. This assumption has been
employed for years in studies of hadron mass modifications in
nucleus-nucleus collisions \cite{cass99}. Since the experimental
production cross sections essentially are proportional to the
available phase space, it is sufficient to shift the threshold in
a $pN$ collision to
\begin{equation}
\sqrt{s_0^*} = \Pi^0_N(p_N) + \Pi^0_\Lambda (p_\Lambda) + \Pi^0_K
(p_K) \label{shift}
\end{equation}
using (\ref{g26}), where the momenta $p_N, p_\Lambda, p_K$ denote
the relative momenta with respect to the nuclear matter rest
frame.

\subsection{Folding model and its limitations}
We briefly recall the assumptions of the folding model, that is
used for the evaluation of momentum-dependent differential
production probabilities from secondary pion-nucleon collisions in
the CBUU approach and is described in more detail for
proton-nucleus reactions in \cite{15,Sib97,paryev}. The underlying
picture is that the proton impinging on a nucleus at a bombarding
energy $T_{lab}> 400$ MeV is producing a  meson $x$ with momentum
$k_x$ only in the first collision due to the available energy in
the reaction. The Lorentz-invariant differential cross section to
produce a meson in a primary proton-nucleon ($pN$) collision  is
given by \cite{Sib97}
\begin{equation}
\left(E_x {d^3 \sigma ^{NN}_x \over d^3k_x}\right)_{prim.} =
 \int d^3p d \omega
 \ \left(E'_x {d^3\sigma ^e_x(\sqrt{s}) \over d^3k'_x}\right)
\  S ({\bf p},\omega),
\label{fold1}
\end{equation}
\noindent where the Pauli-blocking factor for the final nucleon
states is neglected since kinematically the nucleons end up in an
unoccupied regime in momentum space at the bombarding energies of
interest. In eq. (\ref{fold1}) $S({\bf p},\omega)$ stands for the
target spectral function (normalized to 1) which can be taken from
experiment, e.g. Ref. \cite{Sick}, or parametrized accordingly. In
eq. (\ref{fold1}) the primed indices denote meson momenta in the
individual nucleon-nucleon cms frame which have to be
Lorentz-transformed to the detection frame. The quantity
$\sqrt{s}$ is the invariant energy of the individual NN system,
while the elementary differential cross section $d^3\sigma ^e_x
(\sqrt{s})/d^3k_x$ in (\ref{fold1}) is  parame\-trized according
to phase space (cf. \cite{15}) and assumed to be isotropic in the
$NN$ cms frame. The momentum differential production probability
per $NN$ collision is obtained from (\ref{fold1}) when dividing by
the total $NN$ cross section.

In order to obtain the inclusive differential kaon cross section
in a $p + A$  reaction within the folding model one has to
multiply the differential cross section (\ref{fold1}) by the
number of first-chance collisions $N_1(A)$. This number is
approximately given by the area of the target divided by the $p N$
cross section, i.e. $N_1(A) \approx \pi R^2_{target}/ \sigma_{p
N}$. To be more accurate one can use Glauber theory which leads to
$N_1 \approx  7.3$ for p + $^{12}C$ \cite{cass95,debowski}. The
contributions to the $K^+$ yield from secondary or further
sequential $NN$ collisions is discarded in the folding model
at subthreshold energies
due to the energy straggling of the impinging proton and due to
the fact that already the first-chance collisions only give a
minor contribution to the $K^+$ yield observed experimentally.
Note, however, that within the CBUU approach employed here the number of
collisions is calculated dynamically and thus no assumption or
separate model for $N_1(A)$ has to be invoked.

Apart from the primary reaction channels described above, the
first NN collision may also lead to the excitation of a $\Delta$-
resonance or even higher baryon resonances (e.g. N(1440), N(1535)
..) which decay to nucleons and essentially pions due to their
short lifetimes of $\approx $ 1 fm/c or collide with another
nucleon before decaying. The $\Delta N$ or resonance-nucleon
reactions are treated in the CBUU approach dynamically as
described before. Energetic secondary pion-nucleon collisions,
however, suffer from very low statistics in transport simulations
of $p+A$ reactions. On the other hand, their contribution to the
kaon yield is expected to be high at subthreshold energies
\cite{cass95}.  We thus implement the secondary pion-nucleon
production channels for kaons following concepts of the folding
model.

 The differential cross section to produce a $K^+$ meson in a
collision of an on-shell pion with a nucleon from the target at
invariant energy $\sqrt{s}$ then is given by
\begin{equation}
\left(E_K {d^3 \sigma^{\pi N}_K \over d^3k}\right) = \sum_c \int
{d^3p d \omega}
 \ \left(E'_K {d^3\sigma ^{\pi N}_K(\sqrt{s}) \over
 d^3k'}\right)_c
\  S ({\bf p},\omega)
\label{fold1p}
\end{equation}
similar to (\ref{fold1}), where now the differential cross
sections for the reactions $\pi N \rightarrow K^+Y$ enter. Here,
the index $c$ denotes all individual channels while the hyperon
$Y$ stands again for a $\Lambda$ or $\Sigma$ baryon.

In order to evaluate the $K^+$ yield from
secondary $\pi N \rightarrow  K^+ Y$ collisions  in the folding model one folds
the primary pion distribution  -- given by the primary differential pion cross
section that is divided by the total $pN$ cross section $\sigma_{tot}$ -- with
the nucleon spectral function $S({\bf p}, \omega)$  and the
invariant production cross section, $i.e.$
\begin{eqnarray}
&& \left(E_k \frac{d^3\sigma _K}{d^3k}\right)_{\sec .} =  \int
\int \frac{d^3p d\omega}{\sigma_{tot}}  \frac{d^3k'_\pi}{E'_\pi}
 S ({\bf p},\omega) \ g_\pi(A)  \label{fold2}\\
&&\times \left(E'_k {d^3\sigma _{\pi N\rightarrow Y K^+}({\bf
p,k_\pi} ) \over d^3k'} \right) \ \left(\frac{d^3 \sigma_{pN
\rightarrow \pi X}(\sqrt{s'})}{d^3 k'_\pi}\right)_{prim.}.
\nonumber
\end{eqnarray}
\noindent  In Eq. (\ref{fold2}) the single prime indices denote
the system of the intermediate pion and a target nucleon.
Moreover,
\newline $E'_\pi {d^3 \sigma_{pN \rightarrow \pi
X}(\sqrt{s'})}/{d^3 k'_\pi}$ stands for the $\pi$-meson
differential cross section, which is calculated in line with
(\ref{fold1}), while $\sigma_{tot}$ denotes the total
proton-nucleon cross section. The factor $g_\pi(A)$ in
(\ref{fold2}) accounts for the probability that the pion interacts
again with a target nucleon (cf. \cite{Sib97}). Note, that by Eq.
(\ref{fold2}) the intermediate pion is assumed to be on-shell
which might not hold for deep subthreshold kaon production. As in
case of the primary channel the expression (\ref{fold2}) has to be
multiplied by $N_1(A)$ in the folding model. Furthermore, some
estimate for the secondary rescattering probability $g_\pi(A)$ has
to be employed in the folding model as e.g. described in Ref.
\cite{PINOT}. As mentioned before, when using the differential
probabilities in a perturbative transport approach, the pion
rescattering probabilities are calculated dynamically without
approximation on $g_\pi(A)$.

The one- and two-step folding model has been used often in the
analysis of kaon or $\eta$-meson production
\cite{15,cass95,debowski,Sib97,paryev,Paryev2,Paryev3} initially
employing parametrized momentum distributions (cf. Refs.
\cite{15,cass95}) instead of spectral functions. They allow for an
estimate of differential cross sections in case of weakly
interacting probes and may serve as reference calculations for
more involved simulations employing all initial and final state
interactions during the reaction as the CBUU approach employed
here.
\begin{figure}[h]
\centerline{\psfig{figure=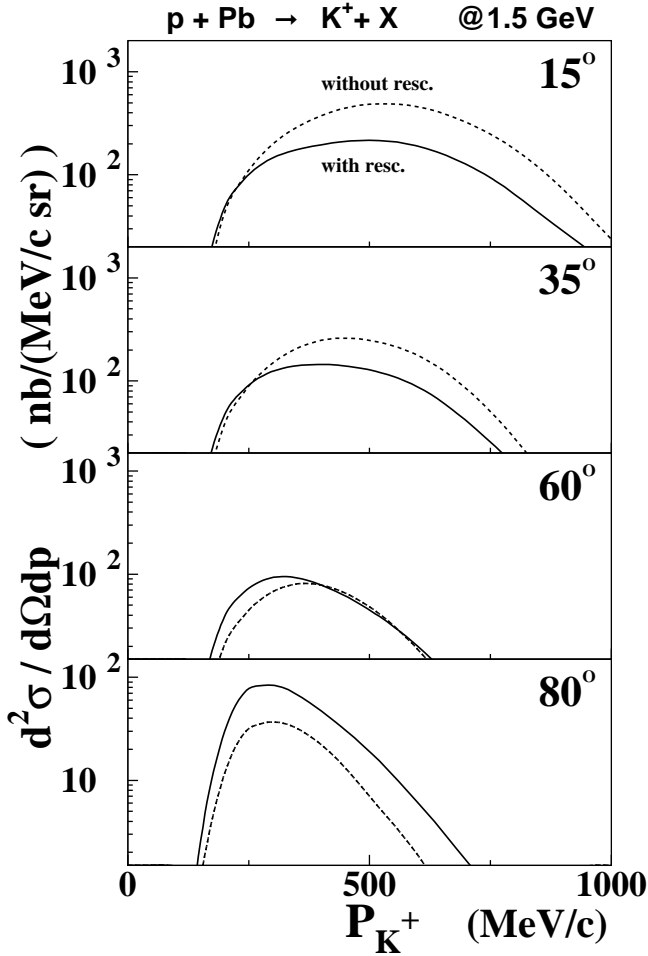,width=8.5cm}}
\caption{Comparison
of the differential $K^+$ spectra for $p+Pb$ at 1.5 GeV at
different angles in the laboratory from 15$^0-80^o$ . The solid lines are obtained
from CBUU calculations including kaon rescattering whereas the
dashed lines are without rescattering.  }
 \label{bild3}
\end{figure}

\subsection{$K^+$ rescattering}
As indicated above, the folding model is useful for the evaluation
of total $K^+$ yields, however, becomes questionable in case of
differential spectra especially for heavy targets like $Au$ or
$Pb$  since kaon elastic rescattering cannot be described in a
straightforward manner. In order to show the influence of $K^+$
rescattering on the kaon spectra at different angles we show in
Fig. \ref{bild3} a comparison of our CBUU calculations with (solid
lines) and without (dashed lines) kaon rescattering for $p+Pb$ at
1.5 GeV. As can be seen from Fig. \ref{bild3} the $K^+$ yield in
forward direction ($\theta \leq 15 ^0$) is  reduced by  up to a
factor of 2, while for large angles in the laboratory (80$^o$) the
kaon spectra become enhanced and also shifted to lower momenta in
the laboratory. Thus rescattering has to be included as a
necessary ingredient for the calculations if comparisons to
differential spectra are made on an absolute scale and especially,
if one attempts to extract kaon potentials from the spectral shape
(see below). We note that kaon rescattering thus will always be
included in the calculations to be shown in the following.
Furthermore, we mention that the results from the folding model
(\ref{fold1}), (\ref{fold2}) agree with the resulting spectra from
the CBUU calculation for $p+Pb$ and $p+C$ at 1.5 GeV within  30\%
when neglecting final state interactions as well as nuclear and
Coulomb potentials.

\section{Comparison to experimental data}
\begin{figure}[h]
\centerline{\psfig{figure=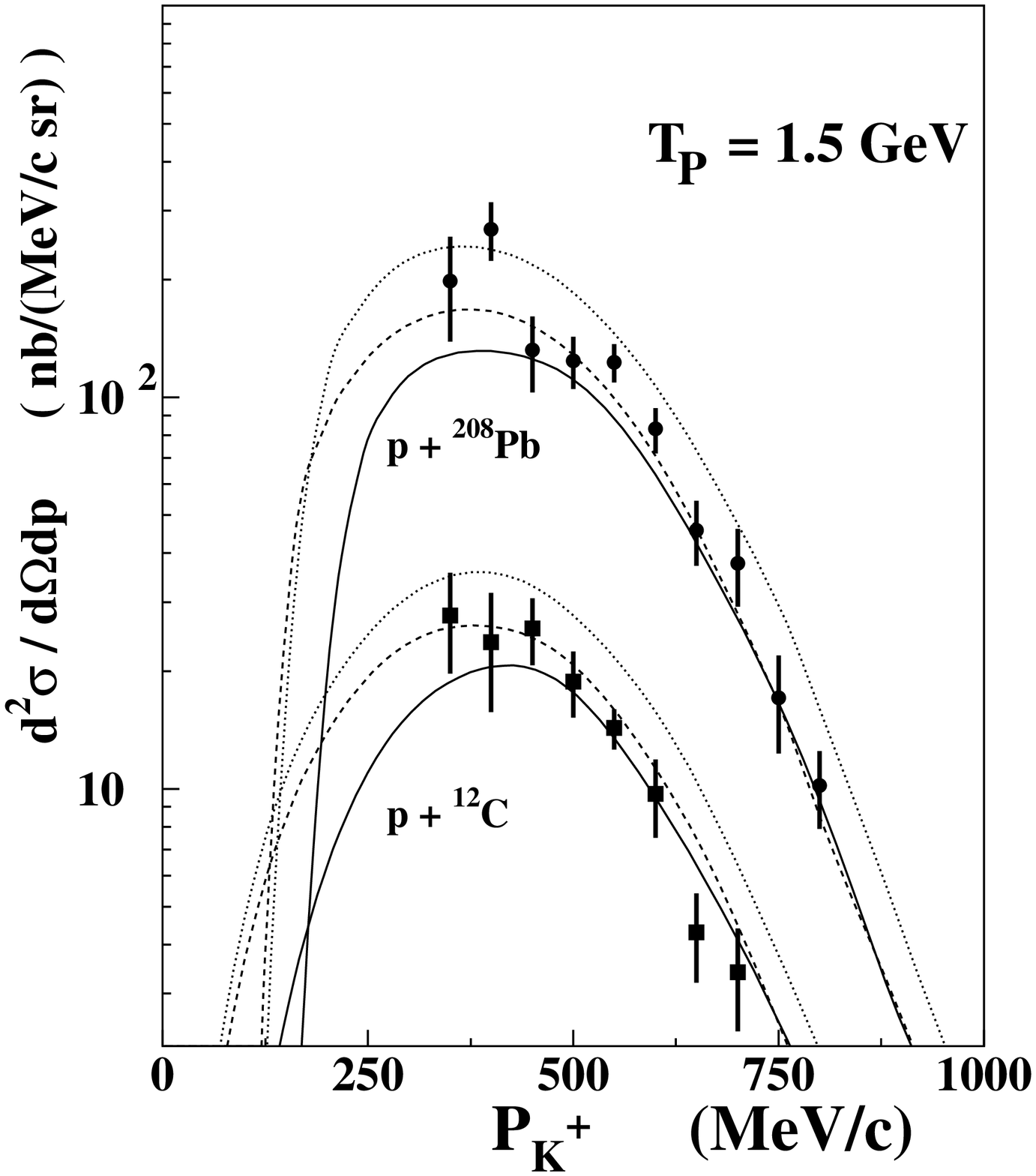,width=8.5cm}}
\caption{Comparison of the CBUU calculations for the differential
$K^+$ spectra for $p+Pb$  and $p+C$  at 1.5 GeV and $\theta_{lab}=
40 \pm$ 5$^0$ with the experimental data from Ref.
\cite{debowski}. The dotted lines are obtained from CBUU
calculations without baryon and kaon potentials, the dashed lines
show the results with baryon potentials included while the solid
lines correspond to calculations including both, nucleon and kaon
potentials. Note, that the effect of the repulsive kaon potential
is a reduction of the total $K^+$ yield as well as a shift of the
spectra at lower momenta. }
 \label{bild4}
\end{figure}

In this Section we compare our calculations to the experimental
$K^+$ spectra available from 1.0 GeV $-$ 2.5 GeV bombarding energy
on different targets.  In order to have an identical assignment of
lines in this section the dotted lines in Figs. \ref{bild4} - 9
correspond to CBUU calculations without baryon and kaon
potentials, the dashed lines show the results with baryon
potentials included while the solid lines reflect calculations
including both, nucleon and kaon potentials as specified in Fig.
\ref{bild1}. In all calculations, however, the Coulomb potential
will be included by default.

We start in Fig. \ref{bild4} with a comparison to the data of the
KaoS/SPES3 Collaboration for the differential $K^+$ spectra for
$p+Pb$ and $p+C$  at 1.5 GeV and $\theta_{lab}= 40 \pm$ 5$^0$ from
SATURNE \cite{debowski}. The experimental spectra for the $Pb$
target are seen to be described roughly within the error bars for
all calculations, i.e. with and without potentials, such that one
is not very sensitive to in-medium potentials at 1.5 GeV in the
momentum range above 350 MeV/c. For lower kaon momenta the
repulsive $K^+$ potential leads to a sizeable decrease (or shift)
in the spectra which can be attributed to the additional
acceleration of the kaons by the nuclear $K^+$ potential when
propagating from the nuclear interior to the vacuum. In case of
the $^{12}C$ target the effects from the momentum-dependent
nucleon potentials as well as from the $K^+$ potential are similar
to the $Pb$ target. It both cases the calculations without
potentials (dotted lines) slightly overestimate the data, but it
is not possible to draw already some conclusion on the actual size
of the $K^+$ potential since a single comparison might suffer from
systematic errors.

\begin{figure}[h]
\centerline{\psfig{figure=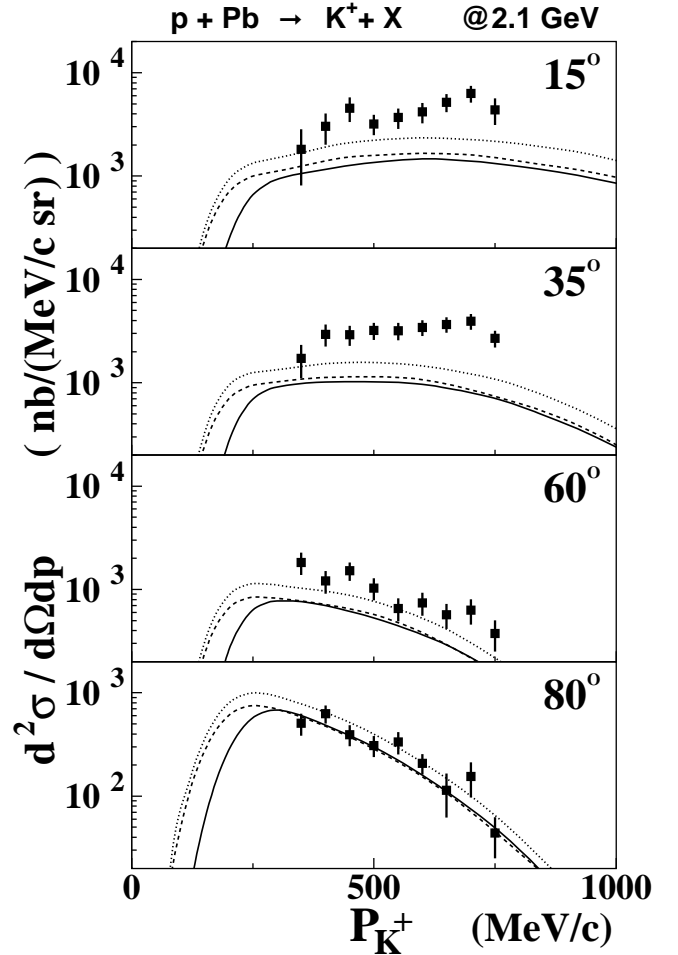,width=8.5cm}}
\caption{Comparison of the CBUU calculations for the differential
$K^+$ spectra for $p+Pb$  at 2.1 GeV  with the experimental data
from Ref. \cite{schnetzer} at different angles in the laboratory.
The dotted lines are obtained from CBUU calculations without
baryon and kaon potentials, the dashed lines show the results with
baryon potentials included while the solid lines correspond to
calculations including both, nucleon and kaon potentials.}
 \label{bild5a}
\end{figure}

\begin{figure}[h]
\centerline{\psfig{figure=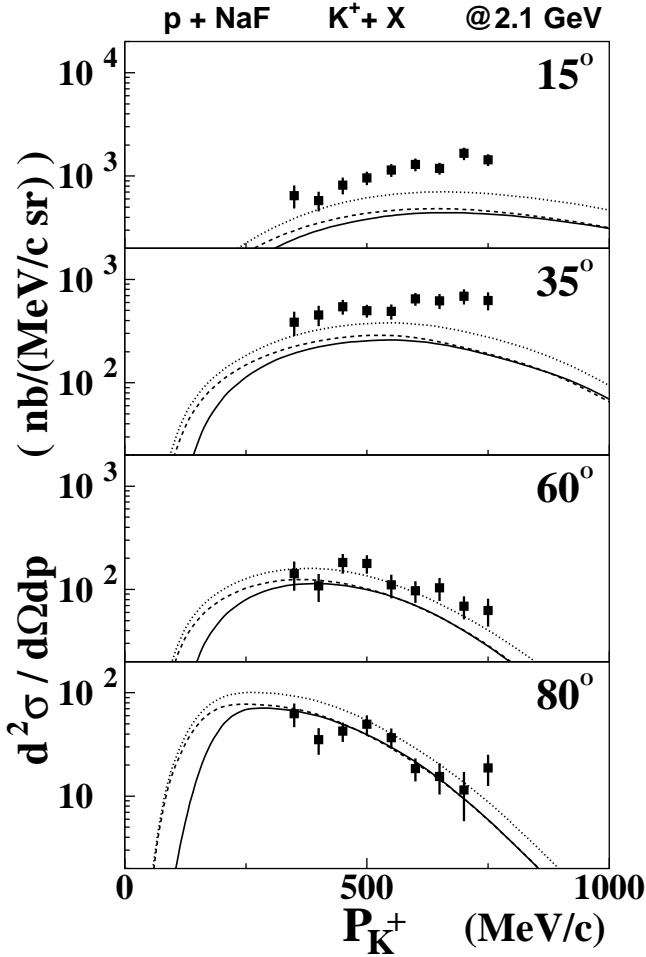,width=8.5cm}}
 \caption{Comparison
of the CBUU calculations for the differential $K^+$ spectra for
 $p+NaF$  at 2.1 GeV  with the
experimental data from Ref. \cite{schnetzer} at different angles
in the laboratory. The dotted lines are obtained from CBUU
calculations without baryon and kaon potentials, the dashed lines
show the results with baryon potentials included while the solid
lines correspond to calculations including both, nucleon and kaon
potentials.}
 \label{bild5b}
\end{figure}

In Figs. \ref{bild5a} and \ref{bild5b} we compare  the CBUU calculations for the
differential $K^+$ spectra for $p+Pb$  and $p+NaF$
 at 2.1 GeV  with the experimental data from the LBL Berkeley
\cite{schnetzer} for different laboratory angles from 15 - 80$^0$.
At this bombarding energy the nucleon and $\Lambda$ final momenta
on average are above 0.6 GeV/c such that their potentials at
finite density (cf. Fig. \ref{bild1}) are repulsive. As a
consequence the calculated kaon yield decreases when including the
baryon potentials in the final state. Taking into account
additionally the repulsive $K^+$ potential decreases essentially
the spectrum for momenta below 250 MeV/c, but leaves the spectrum
practically unmodified for higher momenta since the relative
change of the $K^+$ energy by the kaon potential is only small.

\begin{figure}[h]
\centerline{\psfig{figure=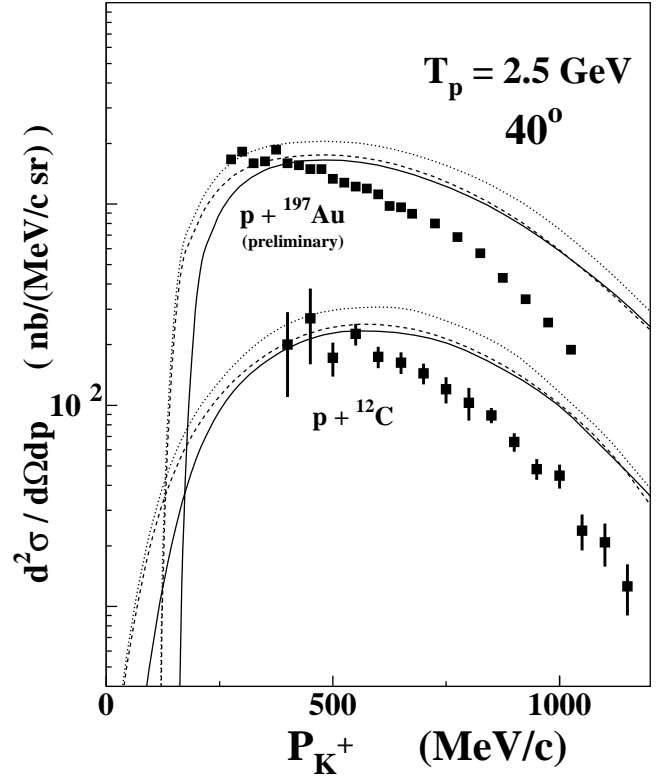,width=8.5cm}}
 \caption{Comparison
of the CBUU calculations for the differential $K^+$ spectra for
$p+Au$ (upper part) and $p+C$ (lower part) at 2.5 GeV  and
$\theta_{lab}= 40 \pm$ 5$^0$ with the experimental data from Refs.
\cite{scheinast,debowski}. The dotted lines are obtained from CBUU
calculations without baryon and kaon potentials, the dashed lines
show the results with baryon potentials included while the solid
lines correspond to calculations including both, nucleon and kaon
potentials.}
 \label{bild6}
\end{figure}

In all approximations the experimental spectra are underestimated
by  factors $\sim 2-3$ at 15$^o$ and 35$^o$, which at first sight
might be attributed to an improper energy dependence of the
calculations. However, a comparison of the CBUU calculations  for
$p+Au$  with the preliminary data from Ref. \cite{scheinast}
(taken at GSI)  and $p+C$  at 2.5 GeV and $\theta_{lab}= 40 \pm$
5$^0$ with the experimental data from SATURNE \cite{debowski} in
Fig. \ref{bild6} shows that these spectra  are overestimated by up
to a factor of 2--3 at higher kaon momenta.
 Note,  that the corresponding
data from Ref. \cite{scheinast} so far have to be considered as
preliminary. These findings suggest that the spectra from Ref.
\cite{schnetzer} are systematically too high by about factors of
$2-3$ or the data from the KaoS Collaboration too low (by about
the same factor).

\begin{figure}[h]
\centerline{\psfig{figure=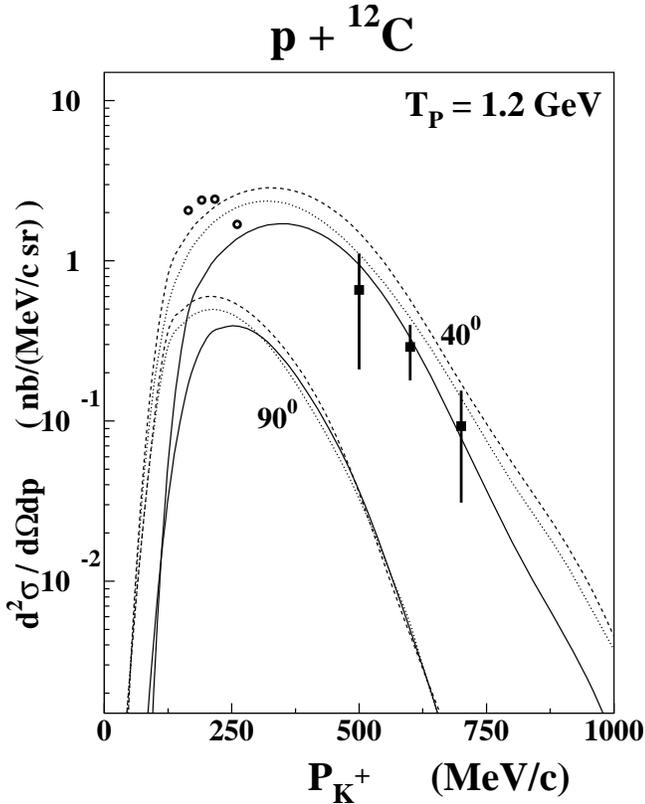,width=8.5cm}}
 \caption{Comparison
of the CBUU calculations for the differential $K^+$ spectra for
$p+C$  1.2 GeV   with the experimental data from Ref.
\cite{badala} at $\theta_{lab}= 90^0 $ (open circles; lower lines)
 and $\theta_{lab}= 40
\pm 5^0$ \cite{debowski} (full squares; upper lines). The dotted
lines are obtained from CBUU calculations without baryon and kaon
potentials, the dashed lines show the results with baryon
potentials included while the solid lines correspond to
calculations including both, nucleon and kaon potentials.}
 \label{bild7}
\end{figure}

To further test the (over-) underprediction of the transport model
we show in Fig. \ref{bild7} a comparison of the CBUU calculations
for the differential $K^+$ spectra for $p+C$ at 1.2 GeV with the
experimental data from Ref. \cite{badala} at $\theta_{lab}= 90^0$
(open circles) taken at CELSIUS and $\theta_{lab}= 40 \pm 5^0$
(full squares) from the KaoS/SPES3 Collaboration \cite{debowski}
taken at SATURNE. In this particular case the spectra from Ref.
\cite{debowski} are slightly overestimated by the calculations
when neglecting the kaon potential while the spectra from Ref.
\cite{badala} at 90$^o$ are underestimated by about a factor of
5-6 when neglecting the repulsive $K^+$ potential and by about an
order of magnitude for the repulsive kaon potential included,
which leads again to a sizeable decrease of the spectra at low
momentum. It is not clear to the authors where such discrepancies
might come from.

The CBUU calculations demonstrate that the kaon spectra at $90^o$
and $40^0$ are slightly enhanced (dashed lines)  when taking the
nucleon potential effects into account. Contrary to the
kinematical situation at $T_{lab}$= 2.1 (or 1.5) GeV the nucleon
and $\Lambda$ final momenta here on average are below 0.6 GeV/c,
where the potentials are attractive (cf. Fig. \ref{bild1}), such
that the $K^+$ production becomes enhanced (dashed lines) relative
to the free case (dotted lines). When including additionally the
overall repulsive kaon potential the $K^+$ spectrum drops again
(solid lines).

\begin{figure}[h]
\centerline{\psfig{figure=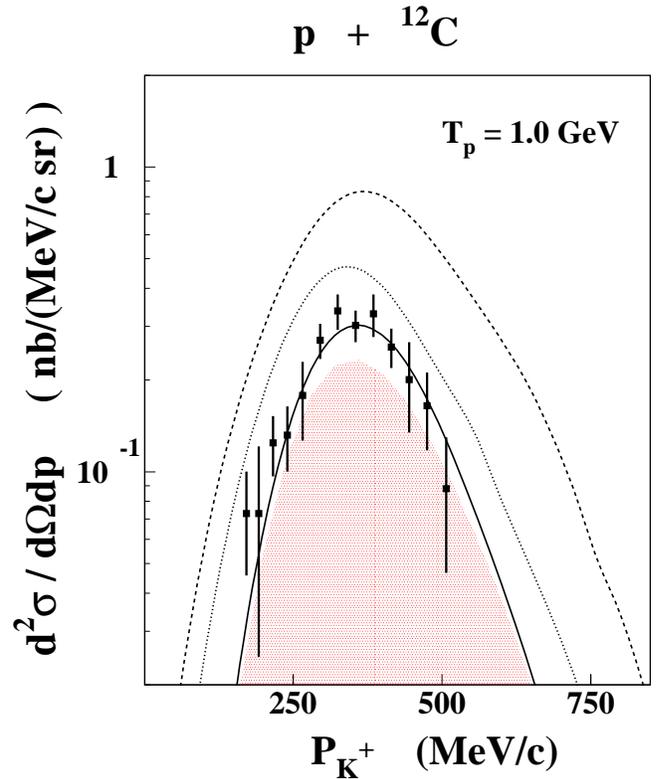,width=8.5cm}}
 \caption{The
calculated differential $K^+$ spectra for $p+C$ at 1.0 GeV for
$\theta_{lab} \leq 12^0$  within the acceptance of the ANKE
spectrometer in comparison to the data from \cite{ANKE}. The
dotted line is obtained from CBUU calculations without baryon and
kaon potentials, the dashed line shows the results with baryon
potentials included while the solid line corresponds to
calculations including both, nucleon and kaon potentials. The
shaded area indicates the contributions from the two-step
mechanisms $\Delta N \rightarrow K^+ Y N$ and $\pi N \rightarrow
K^+ Y$, respectively, for the case of nucleon and kaon potentials,
such that the difference to the solid line corresponds to the
primary $pN$ production channel. }
 \label{bild8}
\end{figure}

We now turn to the kinematical conditions of the ANKE experiments
at COSY-J\"ulich \cite{Barsov}, that have taken  $K^+$ spectra in
forward direction for $\theta_{lab} \leq 12^o$.  The calculated
differential $K^+$ spectra for $p+^{12}C$ at 1.0 GeV
 for $\theta_{lab} \leq 12^0$  are displayed in Fig.
\ref{bild8} in comparison to the data from Ref. \cite{ANKE}. The
dotted lines again are obtained from CBUU calculations without
baryon and kaon potentials, the dashed lines show the results with
baryon potentials included while the solid lines correspond to
calculations with both, nucleon and kaon potentials. At this low
bombarding energy the net attractive baryon potentials in the
final state enhance the $K^+$ yield by about a factor of 2 whereas
the additional repulsive $K^+$ potential leads again to a decrease
by a factor $\sim$ 3. The data from Ref. \cite{ANKE} are rather
well described by the calculations that include the baryon and
$K^+$ potentials (solid line), whereas the other limits clearly
fail. This might be considered as a first indication for the
observation of a repulsive $K^+$ potential in $p+A$ reactions,
however, a full systematics in target mass $A$ and laboratory
energy $T_{lab}$ will be needed to pin down this effect
unambiguously.

The shaded area in Fig. 9 indicates the contributions from the
two-step mechanisms $\Delta N \rightarrow K^+ Y N$ and $\pi N
\rightarrow K^+ Y$, respectively, for the case of nucleon and kaon
potentials included (solid line). Thus the role of secondary
($\Delta$ and pion induced) reaction channels is clearly visible
from Fig. 9 by comparing the shaded area to the solid line, that
correspond to the total spectra. At $T_{lab}$ = 1.0 GeV the
dominant fraction of the $K^+$ yield is due to the secondary
channels in line with the earlier calculations in Refs.
\cite{15,cass95}. Consequently, one does not probe high momentum
components of the nuclear wave function by $K^+$ spectra in a
direct way.

Without explicit representation we mention that at $T_{lab}$ = 2.3
GeV the secondary channels in case of a $Au$ target amount to
about 30\% and in case of a $C$ target to less than 20\%. This
relative change with target mass number is attributed to the fact
that for the small $^{12}C$ target only a fraction of the high
energy pions rescatters in the target and produces $K^+ Y$ pairs.
Moreover, the role of the secondary channels decreases with
increasing kaon momentum such that the high momentum $K^+$ tail of
the spectra is dominated by the first chance $pN$ production
channel as found out before by Paryev \cite{paryev}.

Furthermore, it is worth to point out that the contribution of the
primary channel $pN \rightarrow K^+ Y N$ is enhanced by up to a
factor of 3 within the angular range of $\theta \leq 12^0$ as
compared to the angle integrated yield at the energy $T_{lab}$ =
1.0 GeV. Thus the experimental mass dependence, when expressed in
terms of a scaling $\sim A^\alpha$, does not allow to disentangle
the relative contribution of the different reaction channels in a
satisfying manner for narrow cuts in the $K^+$ angular
distribution.

\begin{figure}[h]
\centerline{\psfig{figure=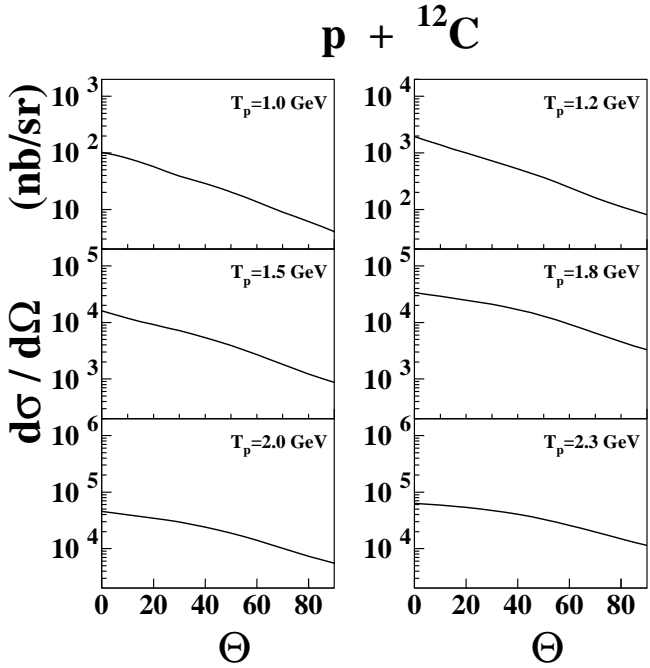,width=8.5cm}}
 \caption{The
calculated angular distribution of the $K^+$ spectra for
$p+C$ at 1.0, 1.2, 1.5, 1.8, 2.0 and 2.3 GeV
for $0.2 \leq p_K \leq 0.5$ GeV/c. The
solid lines are obtained from CBUU calculations including both,
nucleon and kaon potentials.}
 \label{bild11}
\end{figure}

In order to provide some guideline for extrapolations between
experiments measuring $K^+$ spectra at different angles in the
laboratory we show in Fig. \ref{bild11} the angular distribution
of the kaons for momenta 0.2 GeV/c $\leq p_K \leq$ 0.5 GeV/c as
calculated within the CBUU approach for both, baryon and kaon
potentials for $p+ ^{12}C$ at $T_{lab}$ = 1.0, 1.2, 1.5, 1.8, 2.0
and 2.3 GeV. These angular distributions are rather flat within
the angular acceptance of the ANKE spectrometer of $\sim 12^0$,
however, drop substantially for angles larger than 40$^0$. Thus
our calculations (cf. also Fig. 8) do not support the idea of a
'thermal' production mechanism for kaons in case of $p+A$
reactions.

\section{Summary}

In this work we have studied the production of $K^+$  mesons in
proton-nucleus collisions from 1.0 to 2.5 GeV   with respect to
one-step nucleon-nucleon and two-step $\Delta$-nucleon or
pion-nucleon production channels on the basis of a coupled-channel
transport approach (CBUU). We have included the kaon final state
interactions, which are important for heavy targets like $Au$ or
$Pb$, and explored the effects of momentum-dependent potentials
for the nucleon, hyperon and kaon in the nucleus. A comparison of
the transport calculations to the experimental $K^+$ spectra taken
at LBL Berkeley, SATURNE, CELSIUS, GSI and COSY-J\"ulich has shown
that the different data sets are not compatible with each other.
Thus no clear signal on in-medium potentials could be extracted
from our analysis in comparison to experimental data so far.

However, the detailed calculations  demonstrate that precise and
complete spectra show a substantial sensitivity to the potentials
and their momentum dependence. At low bombarding energies of
$\sim$ 1.0 GeV the net attractive potentials for the nucleon and
the $\Lambda$-hyperon in the final state lead to a relative
enhancement of the $K^+$ spectra while at higher bombarding
energies ($\sim$ 2 GeV) the baryon potentials are repulsive and
thus suppress $K^+$ production relative to the free case. This
phenomenon should be seen in the excitation function of the $K^+$
cross section when varying $T_{lab}$ from 1.0 -- 2.5 GeV.
Furthermore, the shape of the spectrum for low $K^+$ momenta in
the laboratory is very sensitive to both, Coulomb and nuclear kaon
potentials, since the kaons are accelerated by both forces when
leaving the nuclear environment and propagating to the continuum.
The relative strength of this momentum shift in the forward $K^+$
spectra is proportional to the square root of the sum of both
potentials, i.e. $\Delta p \approx \sqrt{2 M_K (U_{Coul}+U_K)}$ .
Thus the $K^+$ spectral shape at low momenta (or kinetic energies
$T_K$) allows to determine the strength of the $K^+$ potential
from experimental data in an almost model independent way
especially when comparing kaon spectra from light and heavy
targets at the same bombarding energy \cite{Barsov2} as a function
of $T_K$. Since most of the $K^+$ spectra measured so far have
been taken at higher momenta in the laboratory (except for Ref.
\cite{ANKE}) this finding opens up interesting perspectives for
the ANKE Collaboration at COSY-J\"ulich, which has performed a
systematic study of $K^+$ production in $p+A$ reactions down to
momenta of 150 MeV/c in the laboratory or $T_K \approx $ 23 MeV,
respectively.

\vspace{0.5cm} The authors like to acknowledge valuable
discussions with  M. B\"uscher, V. Koptev, M. Nekipelov, E. Ya.
Paryev, P. Senger, A. Sibirtsev, H. Str\"oher and C. Wilkin on
various issues of this study. Financial support has been provided
by the German BMBF under grant WTZ-POL 99/001 and the Polish State
Committee for Scientific Research under grant 2 P03B 101 19.

\end{document}